August 17, 2018

# Density Matrix Renormalization Group Pair-Density Functional Theory (DMRG-PDFT): Singlet-Triplet Gaps in Polyacenes and Polyacetylenes

Prachi Sharma,[a†] Varinia Bernales,[a†] Stefan Knecht,[b*] Donald G. Truhlar,[a*] and Laura Gagliardi [a*]

[a] *Department of Chemistry, Chemical Theory Center, and Minnesota Supercomputing Institute, University of Minnesota, Minneapolis, Minnesota 55455, United States*

[b] *Laboratory of Physical Chemistry, ETH Zürich, Vladimir-Prelog-Weg 2, CH-8093 Zürich, Switzerland*

[†] These authors contributed equally.

* Corresponding authors

## Abstract

The density matrix renormalization group (DMRG) is a powerful method to treat static correlation. Here we present an inexpensive way to add additional dynamic correlation energy to a DMRG self-consistent field (DMRG) wave function using pair-density functional theory (PDFT). We applied this new approach, called DMRG-PDFT, to study singlet-triplet gaps in polyacenes and polyacetylenes that require active spaces larger than the feasibility limit of the conventional complete active-space self-consistent field (CASSCF) method. The results match reasonably well the most reliable literature values and have only a moderate dependence on the compression of the initial DMRG wave function. Furthermore, DMRG-PDFT is significantly less expensive than other commonly applied ways of adding additional correlation to DMRG, such as DMRG followed by multireference perturbation theory or multireference configuration interaction.

## 1 Introduction

The accurate yet affordable treatment of large molecular systems with close-lying electronic states has long been a target for the development of new quantum chemical methods.[1-7] Molecular systems with degenerate or nearly degenerate electronic states are called inherently multiconfigurational systems or strongly-correlated systems, the correct treatment of which requires a better starting point than the usual independent-electron approach.[8] Electron correlation is sometimes classified into two categories: static and dynamic correlation.[4-6] Static correlation arises when two or more electronic states are close in energy. Systems with high static correlation are called inherently multiconfigurational, strongly correlated, or multireference systems. Dynamic correlation, on the other hand, arises mainly from minimizing short-range repulsion or maximizing long-range dispersion and middle-range dispersion-like interactions of electrons.

Multiconfiguration self-consistent-field (MCSCF) methods, like the complete active space self-consistent field (CASSCF)[9-11] method, where the wave function corresponds to full configuration interaction (FCI) of an active set of electrons in a set of active orbitals, were originally developed to treat static correlation. However, CASSCF is limited by the exponential increase in cost with size of the active space such that 20 electrons in 20 orbitals (20, 20) is the largest reported active space treated by the conventional CASSCF method.[12] The restricted active-space SCF (RASSCF)[13], the generalized active-space SCF (GASSCF),[14] and the occupation restricted multiple active space (ORMAS)[15] approaches can be used to define less-than-full configuration interaction (CI) lists  that make even larger active spaces feasible. For example, in RAS and GAS methods, the active space is further divided into smaller subspaces and only certain excitations are allowed within them, thereby making these methods more affordable. One difficulty with these approaches is to choose the appropriate subspaces and appropriate restrictions on the kinds of included electron excitations because such choices are not systematic and often require subjective chemical intuition. Another difficulty, which applies to all active space methods, is their slow convergence with respect to the size of the active space. The FCI problem converges to the exact solution of the Schrödinger equation as the number *L* of active orbitals is increased, but the convergence is very slow. Thus FCI is not an efficient way to include all the electron correlation. The CASSCF approach was originally designed as a way to recover the static correlation due to nearly degenerate configurations. Inevitably it also includes some of the remaining electron correlation, which is called dynamic correlation, but usually only a small fraction of it. The dynamic correlation not included in a CASSCF calculation is called external correlation, and one of the main challenges in quantum chemistry is to design methods to include the external correlation energy efficiently.

The density matrix renormalization group (DMRG)[16,17] algorithm is a replacement of the FCI solver in the CASSCF or CASCI methods.[18,19] (CASCI is like CASSCF, but the orbitals are not self-consistently optimized for a given configuration list.) The major advantage of DMRG comes from its polynomial scaling with respect to the size of the active space; this allows practical computation of numerically exact solutions for active spaces three to four times larger than standard CASSCF.[20] In DMRG, one may use self-consistently optimized orbitals, as in CASSCF, or pre-determined orbitals, as in CASCI. Although, multiconfigurational methods such as CASSCF or DMRG can be effectively applied to capture static correlation, and they reduce the choosing which orbitals to include in the active space (because they allow more orbitals to be included), they are not efficient at capturing all the correlation energy because of the slow convergence.

A number of approaches have been advanced to make it more efficient to treat the dynamical correlation not included in a MCSCF calculation. For example, approaches that have been combined with DMRG include internally contracted multireference CI (MRCI),[21,22] multireference perturbation theory (MRPT),[23-34] and wave function theory–short range density functional theory (WFT-srDFT).[35]. In the present study we introduce the use of pair-density functional theory (PDFT)[36,37] to include dynamic electron correlation beyond that captured within a DMRG wave function. In the present work we use self-consistently optimized orbitals. The major advantage of DMRG-PDFT is that it treats both static and dynamic electron-correlation and gives accurate results at significantly lower memory and computational costs than multireference perturbation treatments and MRCI approaches.

One example of systems requiring large active spaces is the family of polyacenes (also called acenes) and polyacetylenes, both containing alternating single and double bonds, as shown in Fig. 1. (Note that we only considered linear polyacenes in this article.) Electron correlation in these systems increases with the number of conjugated double bonds,[38-40] and large active spaces are needed, making this a critical test of whether DMRG-PDFT can provide useful accuracy on challenging systems. The acene systems are especially interesting because their electronic properties make them useful for biodegradable and low-cost electronics.[41-44] The smaller acenes, such as naphthalene and anthracene, are used industrially to make various dyes, while larger acenes, like tetracene and pentacene along with their derivatives, are used as organic light-emitting diodes (OLEDs) and organic field-effect transistors (OFETs).[45-56] Recently, tetracene and pentacene have been found to undergo singlet-fission where a high-energy singlet exciton converts to two low-energy triplet excitons, thereby increasing the efficiency of solar cells by up to 40%.[57-64] Polyacetylenes show high electrical conductivity, which further increases on doping with p-type dopants such as $Br_2$, $I_2$, $Cl_2$, and $AsF_5$ .[65-69]

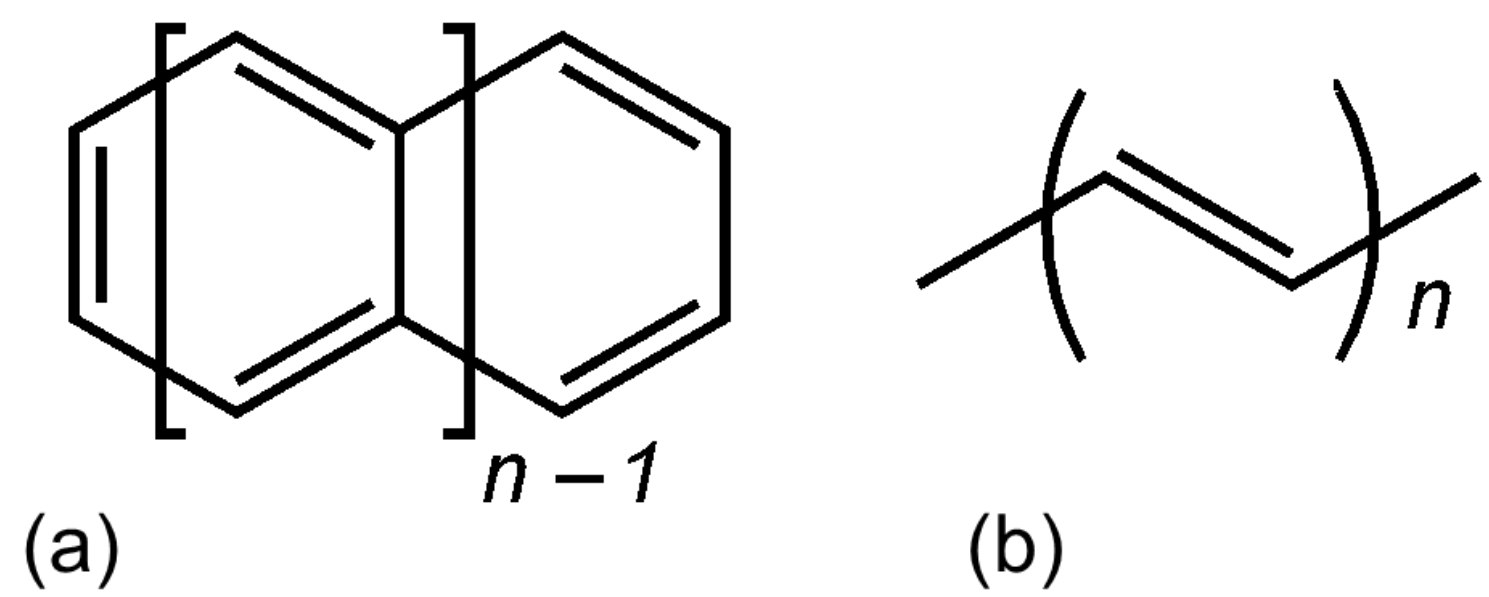

**Fig. 1** (a) *n*-acenes, (b) *n*-polyacetylene

We briefly review DMRG and MC- PDFT in section 2, followed by a presentation of DMRG-PDFT. The computational methods are in section 3. Results and discussion are in section 4, and conclusions are in section 5.

## 2 Theory

In this section, we review the matrix product state (MPS) and DMRG algorithms, and we provide a brief discussion of MC-PDFT, von Neumann entropy, and correlation measures.

## 2.1. Matrix product states (MPS) and density matrix renormalization group (DMRG)

Because DMRG is well described in the literature,[18,19,70-73] we summarize only enough details as are necessary to specify the present applications.

The non-relativistic electronic Hamiltonian may be written in second quantization as[74]

$$\hat{H} = \sum_{pq} h_{pq} E_{pq} + \frac{1}{2} \sum_{pqrs} g_{pqrs} e_{pqrs} + h_{nuc} \qquad (1)$$

where $p$, $q$, $r$, and $s$ denote molecular orbitals, and $h_{pq}$ and $g_{pqrs}$ are one- and two-electron integrals, respectively, $E_{pq}$ is the singlet excitation operator, and $e_{pqrs}$ is the two-electron excitation operator. The excitation operators can be written in terms of creation and annihilation operators as

$$E_{pq} = a^{\dagger}_{p\alpha} a_{q\alpha} + a^{\dagger}_{p\beta} a_{q\beta} \ ,$$

and

$$e_{pqrs} = \sum_{\sigma\tau} a^{\dagger}_{p\sigma} a^{\dagger}_{r\tau} a_{s\tau} a_{q\sigma}, \qquad (2)$$

where Greek letters label the spin functions associated with the molecular orbitals $p$, $q$, $r$, and $s$. The electronic energy for eigenstate $|\Psi\rangle$ is expressed as

$$E = \langle\Psi|\hat{H}|\Psi\rangle = \sum_{pq} h_{pq} D_{pq} + \frac{1}{2} \sum_{pqrs} g_{pqrs} d_{pqrs} + V_{NN} \ , \qquad (3)$$

where $V_{NN}$ is the nuclear-nuclear repulsion energy; $D_{pq}$ and $d_{pqrs}$ are the elements of one-electron and two-electron reduced density matrices (RDMs), respectively.

Each spatial orbital (also referred to as a *site* in DMRG terminology) is associated with four possible occupations:

$$n_r = \{|vac\rangle, |\uparrow\rangle, |\downarrow\rangle, |\uparrow\downarrow\rangle\}. \qquad (4)$$

The eigenstate of the Hamiltonian (1) can be written in an occupation number basis as

$$|\Psi\rangle = \sum_{\{n_j\}} C^{n_1 n_2 \dots n_L} |n_1 n_2 \dots n_L\rangle, \quad (5)$$

where $|n_1 n_2 \dots n_L\rangle$ is an occupation-number vector, which is a particular way of writing a Slater determinant; $L$ is the number of orbitals; and $C^{n_1 n_2 \dots n_L}$ is a coefficient interaction (CI) coefficient, which may be considered to be a tensor of order $L$ with $4^L$ elements. The CI coefficients $C^{n_1 n_2 \dots n_L}$ can also be written as the product of $L$ matrices, labeled $i$ = 1, 2, … $L$, each having $\min(4^i, 4^{L+1-i})$ elements. This is called a matrix product state (MPS).

The size of the matrices increases exponentially to a very large maximum in the middle of the product, then decreases again.[19] If no truncation is made, the MPS is equivalent to full configuration interaction (FCI). For practical work, one retains at most *M* terms in each of the matrix multiplication steps; this approximation is called compression, and *M* is called the bond dimension. Compression is the main approximation of DMRG. The compressed MPS is then variationally optimized to give an upper bound for the ground-state energy. Thus, DMRG may be considered to be a way to calculate an approximate wave function that gives an upper bound to the FCI energy.

In practice one makes another simplification in active-space approaches such as DMRG and CASSCF. Some orbitals are restricted to be doubly occupied in all configuration state functions; these are called inactive orbitals. The orbitals whose occupations are variable are called active orbitals. If one did not truncate the bond dimension in DMRG, the resulting wave function would be the CASSCF wave function, which corresponds to FCI among the active orbitals. In this context, DMRG may be considered to offer a possibility to calculate an approximate CASSCF wave function in a controlled manner such that it gives a variational upper bound to the CASSCF energy. However, its efficiency allows one to use many more active orbitals than in conventional CASSCF. The approximate solution to the large-active space CASSCF problem will typically have a lower (and hence more accurate) energy than the uncompressed solution to the small-active-space CASSCF problem.

### 2.2. Multiconfiguration pair-density functional theory (MC-PDFT)

To correct the DMRG energy for dynamic correlation, we use pair-density functional theory.[36,37] The MC-PDFT energy, for a generic multiconfiguration (MC) wave function is expressed as

$$E^{\text{MC-PDFT}} = \sum_{\text{pq}} h_{pq} D_{pq} + \frac{1}{2}\sum_{\text{pqrs}} g_{pqrs} D_{pq} D_{rs} + E_{ot}(\rho, \Pi) + V_{NN}. \tag{6}$$

where $E_{ot}$ is the on-top density functional. In MC-PDFT, the kinetic energy and the density needed to calculate the nuclei-electrons interaction energy and classical Coulomb energy of the electron distribution are obtained by the DMRG method. The remaining part of the energy is computed by the on-top functional, which is a functional of the density ($\rho$) and the on-top pair density ($\Pi$).

## 3 Implementation and Computational Methods

The DMRG-PDFT implementation is based on an interface between the existing MC-PDFT code in the *OpenMolcas* 8.3 software package[75,76] and the DMRG code in the *QC-MAQUIS* program[72,77-79].

The MC-PDFT energy can be written in terms of inactive and active orbitals as

$$E^{\text{MC-PDFT}} = 2\sum_i h_{ii} + \sum_{ij}(2g_{iijj} - g_{ijij}) + \sum_{uv} h_{uv} D_{uv} + \sum_{iuv}(2g_{iiuv} - g_{iviu}) + \frac{1}{2}\sum_{uvxy} g_{uvxy} d_{uvxy} + E_{ot}(\rho, \Pi) + V_{NN}, \tag{7}$$

where $i$ and $j$ are inactive and $u, v, x$ and $y$ are active molecular orbital indices, and $\rho$ and $\Pi$ are the density and the on-top density. To calculate the energy according to equation (7), we need the one-electron and two-electron RDMs, from which we calculate the density and the on-top pair-density. The one and two-electron RDMs are obtained from the DMRG wave function, while the density and on-top pair density are calculated according to:

$$\rho = \sum_{ij} D_{ii}\phi_i\phi_i + \sum_{uv} D_{uv}\phi_u\phi_v \tag{8}$$

$$\Pi = \sum_{ij} D_{ii}D_{jj}\phi_i\phi_j\phi_i\phi_j + \sum_{iiuv} D_{ii}\, D_{uv}\phi_i\phi_i\phi_u\phi_v + \sum_{uv} d_{uvxy}\, \phi_u\phi_v\phi\Phi_x\phi_y \tag{9}$$

where again $i$ and $j$ are inactive and $u, v, x$ and $y$ are active molecular orbital indices.

The vertical singlet–triplet gaps for polyacetylenes and polyacenes were computed using DMRG-PDFT. The geometries of the polyacenes for both the singlet ($1^1A_g$) and triplet ($1^3B_{3u}$) states were taken from ref 38. We optimized  the geometry of polyacetylenes at the same level of theory as the polyacenes, using the B3LYP exchange–correlation functional[80,81] and the 6-31G(d,p) basis set. The optimized structures were then used to perform DMRG calculations followed by MC-PDFT calculations. All MC-PDFT calculations were performed with the tPBE on-top density functional.[82] The 6-31+G(d,p) basis set was used for all the calculations.

The active space is denoted as usual as (*n*, *m*) where *n* is the number of active electrons, and *m* is the number of active orbitals. The active spaces used here correspond to all the π electrons distributed in all the valence π orbitals; thus *n* = *m*. For all polyacene calculations, we constrained the wave functions to $D_{2h}$ symmetry, where the ground state has symmetry $^1A_g$, and the lowest triplet state has symmetry $^3B_{3u}$. The convergence with respect to the bond dimension *M* was tested from 100 to 2000 for naphthalene and anthracene, and *M* was set to 500 for higher acenes. For the polyacetylene calculations, $C_{2h}$ symmetry was imposed. The ground state has symmetry $^1A_g$, and the lowest triplet state has symmetry $^3B_u$.

The DMRG calculations were performed with QCMAQUIS.[72,77-79,83]

## 4 Results and Discussion

### 4.1. Singlet–Triplet gap in polyacenes

The results for vertical and adiabatic singlet-triplet (S-T) gaps of naphthalene are reported in Table 1 as functions of bond dimension *M*. Convergence is reached at *M* = 200 with a value for the DMRG vertical excitation energy of 3.05 eV. The DMRG-PDFT vertical excitation energy is 3.35 eV, with a deviation of 1.5% from CCSD(T)/CBS values reported in ref [84]; in contrast plain DMRG has a 7.4% deviation. The DMRG-PDFT adiabatic excitation energy is 2.91 eV, which is within the range of values (2.65–2.92 eV) calculated by various methods in the refs. 84,85, 86 and 87.

**Table 1** Vertical and adiabatic singlet–triplet gap ($E_{triplet} - E_{singlet}$, in eV) for naphthalene and convergence with *respect to M*

| *M* | **DMRG** | | **DMRG-PDFT** | | **GAS-PDFT**[a] | | **Other literature values** | |
|---|---|---|---|---|---|---|---|---|
| | Vert. | Ad. | Vert. | Ad. | Vert. | Ad. | Vert.[b] | Ad.[c] |
| | | | | | 3.36 | 3.06 | 3.30 | 2.79 |
| 100 | 3.08 | 2.67 | 3.31 | 2.89 | | | | |
| 200 | 3.05 | 2.66 | 3.35 | 2.91 | | | | |
| 500 | 3.05 | 2.66 | 3.35 | 2.91 | | | | |
| 1000 | 3.05 | 2.66 | 3.35 | 2.91 | | | | |

[a] from ref 38.
[b] CCSD(T)/CBS from ref. 84, where “CBS” denotes extrapolation to a complete one-electron basis set.
[c] Average of four values in in the range 2.65–2.92 eV as calculated in refs. 84, 85, 86 and 87.

In Table 2, vertical and adiabatic singlet-triplet (S-T) gaps for different bond dimensions are reported for anthracene. We see convergence at *M* = 1000, and we observe a significant improvement in both vertical and adiabatic singlet–triplet gaps when the MC-PDFT method is used to add extra correlation energy to the DMRG wave function. The S-T DMRG predicted gap decreases when more degrees of freedom are considered (i.e., when the bond dimension is increased), but the DMRG-PDFT gap increases along the same sequence of *M* values. The DMRG-PDFT results show less dependence on *M* than do the plain DMRG results; in fact DMRG-PDFT gives results close to the literature values[38,84-87,88] even for an M value as small as *M* =100.

**Table 2** Vertical and adiabatic singlet–triplet gaps ($E_{triplet} – E_{singlet}$, in eV) for anthracene and convergence with respect to *M*

| *M* | **DMRG** | | **DMRG-PDFT** | | **GAS-PDFT**[a] | | **Other literature values** | |
|---|---|---|---|---|---|---|---|---|
| | Vert. | Ad. | Vert. | Ad. | Vert. | Ad. | Vert.[b] | Ad.[c] |
| | | | | | 2.22 | 1.97 | 2.46 | 2.00 |
| 100 | 2.46 | 2.05 | 2.28 | 2.00 | | | | |
| 200 | 2.42 | 2.03 | 2.26 | 1.97 | | | | |
| 500 | 2.34 | 1.97 | 2.33 | 2.00 | | | | |
| 1000 | 2.31 | 1.96 | 2.36 | 2.02 | | | | |
| 2000 | 2.30 | 1.95 | 2.38 | 2.04 | | | | |

[a] from ref 38.
[b] CCSD(T)/CBS from ref. 84, where "CBS" denotes extrapolation to a complete one-electron basis set.
[c] Average of the values in refs. 84, 85, 86, 87 and 88.

The vertical and adiabatic S-T gaps for various systems, from naphthalene to heptacene, obtained with *M* = 500 are compared to GAS-PDFT and to other literature values in Table 3. We observe that DMRG-PDFT gives very good agreement with the reference data for adiabatic singlet–triplet gaps with a mean unsigned deviation (MUD) of only 0.07 eV while for vertical excitation, it gives an MUD of 0.16 eV. Furthermore, our results are in good agreement with the GAS-PDFT excitation values.

The adiabatic singlet–triplet gaps for DMRG and DMRG-PDFT are compared with literature values in Fig. 2, showing reasonable agreement for DMRG and good agreement for DMRG-PDFT.

**Table 3** Vertical and adiabatic singlet–triplet gaps ($E_{triplet} – E_{singlet}$, in eV) for polyacenes

| | **DMRG**[a] | | **DMRG-PDFT**[a] | | **GAS-PDFT**[b] | | **Other literature values**[c] | |
|---|---|---|---|---|---|---|---|---|
| | Vert. | Ad. | Vert. | Ad. | Vert. | Ad. | Vert. | Ad. |
| **Naphthalene (10, 10)** | 3.05 | 2.66 | 3.35 | 2.91 | 3.36 | 3.06 | 3.30 | 2.79 |
| **Anthracene (14, 14)** | 2.34 | 1.97 | 2.32 | 2.00 | 2.22 | 1.97 | 2.46 | 2.00 |
| **Tetracene (18, 18)** | 1.88 | 1.54 | 1.58 | 1.37 | 1.69 | 1.46 | 1.75 | 1.48 |
| **Pentacene (22, 22)** | 1.56 | 1.24 | 1.13 | 0.98 | 1.29 | 1.10 | 1.36 | 1.05 |
| **Hexacene (26, 26)** | 1.19 | 0.93 | 0.79 | 0.73 | 0.99 | 0.85 | 0.99 | 0.81 |
| **Heptacene (30, 30)** | 0.81 | 0.67 | 0.61 | 0.62 | 0.75 | 0.72 | 0.78 | 0.60 |
| **MUD**[d] | 0.16 | 0.10 | 0.16 | 0.07 | 0.08 | 0.09 | | |

[a] $M = 500$. [b] from ref. 38. [c] Average of the available values from refs 84-88 as summarized in Table 1 of ref. 38. [d] Mean unsigned deviation from values in last two columns

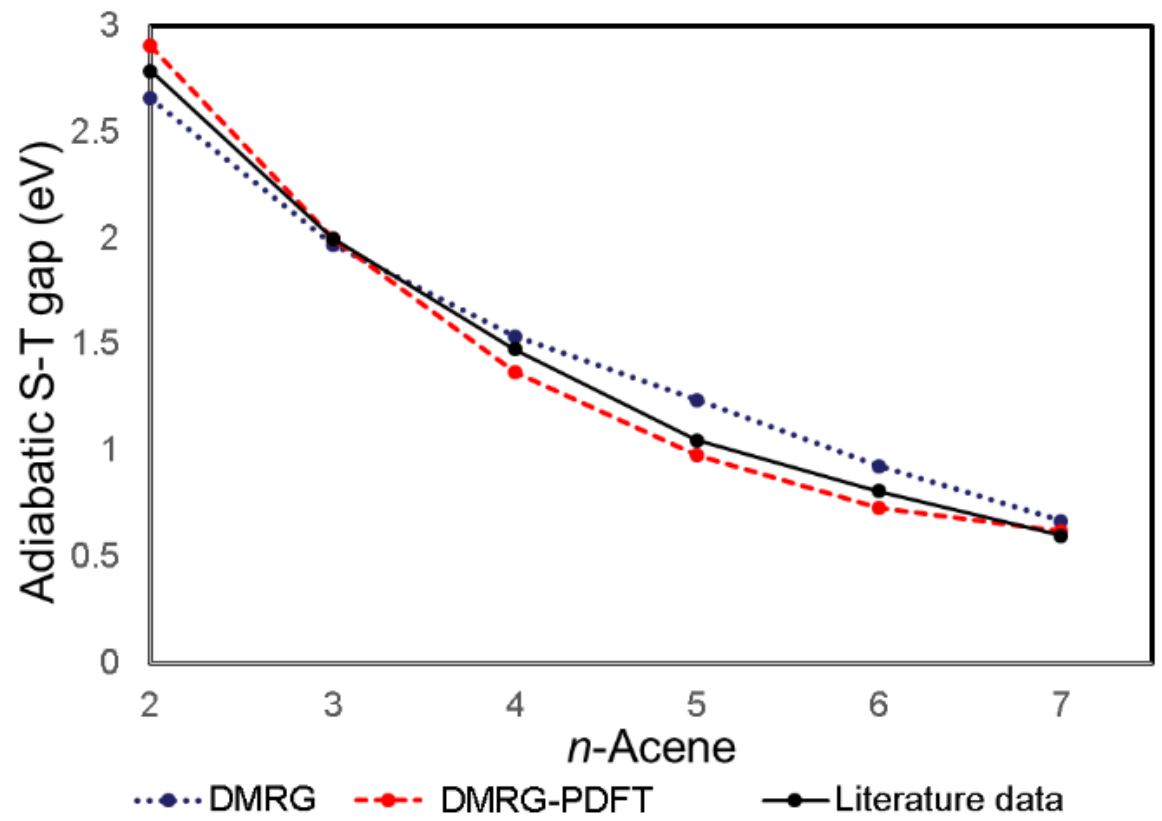

**Fig. 2** Adiabatic singlet–triplet gaps for polyacenes. Literature values from last two columns of Table 3.

### 4.2. Singlet–triplet gap in polyacetylenes

Singlet–triplet energies for polyacetylenes as calculated by DMRG and DMRG-PDFT are reported in Table 4. For smaller polyacetylenes (ethylene to octatetraene), DMRG-PDFT results agree well with the experimental values with an MUD of 0.18 eV. Our DMRG-PDFT results agree with CASPT2 results within 0.2 eV. Note that we used a small bond dimension of 50 for all the acetylene singlet–triplet gaps presented in Table 5. For acetylenes smaller than hexatriene, the DMRG results converge to the CASSCF values, which is not surprising because we found that ethylene and butadiene calculations need *M* values of only 4 and 16 to reproduce the CASSCF results. However, as could be expected, the deviation between the DMRG and CASSCF values increases for larger acetylenes.

**Table 4** Vertical singlet–triplet gap ($E_{triplet} – E_{singlet}$, in eV) for polyacetylenes

| number of monomers | Active space | DMRG | DMRG-PDFT | CASSCF | CASPT2 | Literature values | Exp.[a] |
|---|---|---|---|---|---|---|---|
| **1** | (2,2) | 4.34 | 4.67 | 4.34 | 4.54 | 4.63 [b] | 4.3-4.6 |
| **2** | (4,4) | 3.37 | 3.46 | 3.37 | 3.38 | 3.45 [b],3.20[d] | 3.22 |
| **3** | (6,6) | 2.80 | 2.79 | 2.80 | 2.73 | 2.80 [b], 2.40 [d] | 2.61 |
| **4** | (8,8) | 2.43 | 2.37 | 2.43 | 2.33 | 2.42 [c], 2.10[d] | 2.10 |
| **5** | (10,10) | 2.29 | 1.99 | 2.19 | 2.07 | 2.20[c],1.89 [d] | |
| **6** | (12,12) | 2.20 | 1.79 | 2.01 | 1.88 | 2.00 [c] | |
| **7** | (14,14) | 2.17 | 1.59 | 1.88 | 1.75 | 1.90[c] | |
| **8** | (16,16) | 2.20 | 1.52 | | | | |
| **9** | (18, 18) | 1.03 | 0.07 | | | | |
| **MUD** [e] | | 0.20 | 0.18 | 0.20 | 0.16 | | |

[a] Experimental band maxima for ethylene[89-92], butadiene[93], and hexatriene[94].
[b] CCSD(T)/ cc-pVTZ result from ref. 95.
[c] UCCSD result from ref. 96.
[d] Multireference Møller-Plesset study corrected for basis-set and active-space effects, from ref. 97.
[e] Mean unsigned deviation from experiment

We plotted the convergence of singlet–triplet gaps in DMRG and DMRG-PDFT with respect to conventional CASSCF and MC-PDFT in Fig. 3. DMRG-PDFT converges faster with respect to *M* and shows less dependence on *M* than does DMRG. In Fig. 4, we show the average time required for DMRG, DMRG-PDFT ($M = 50$), and CASPT2 calculations for polyacetylenes. All the calculations were performed on a single processor with a maximum memory of 90 gigabytes. The DMRG-PDFT calculations take considerably less time than the corresponding CASPT2 calculations for higher polyacetylenes (five or more monomers). For example, for seven monomers, CASPT2 takes fifteen times longer than DMRG-PDFT.

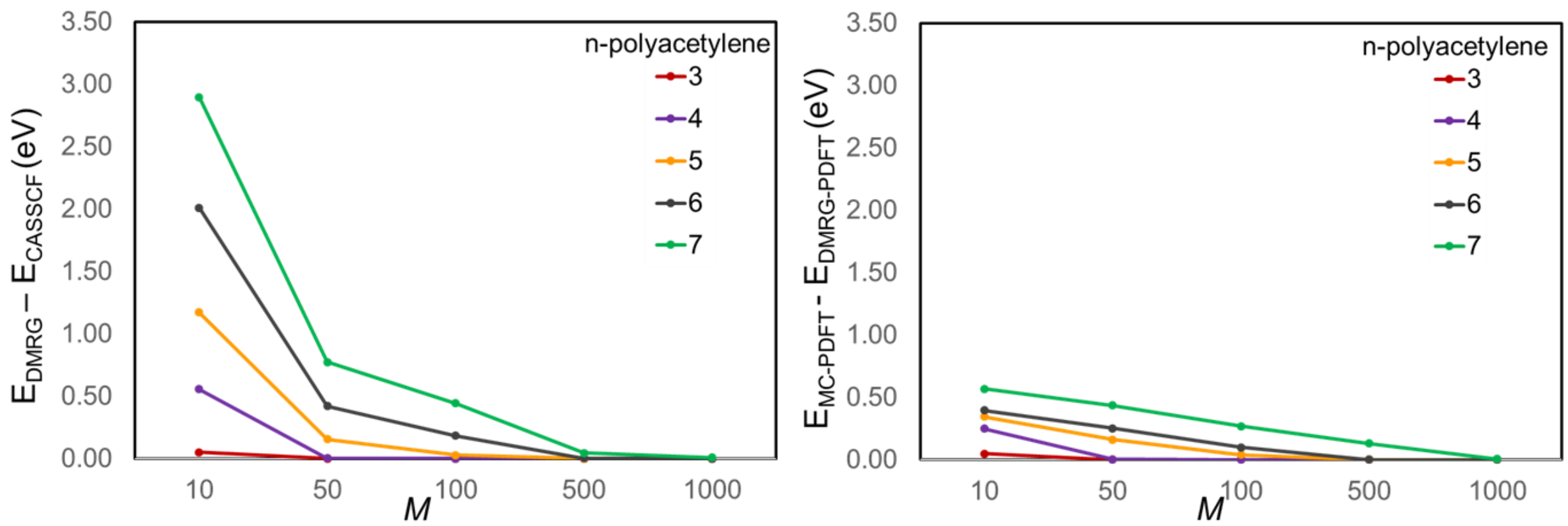

**Fig. 3**. Convergence of (a) DMRG and (b) DMRG-PDFT with respect to conventional CASSCF and MC-PDFT, respectively. The ordinate is the mean of the difference between (a) conventional CASSCF and DMRG energies and (b) conventional MC-PDFT and DMRG-PDFT energies for singlet and triplet states.

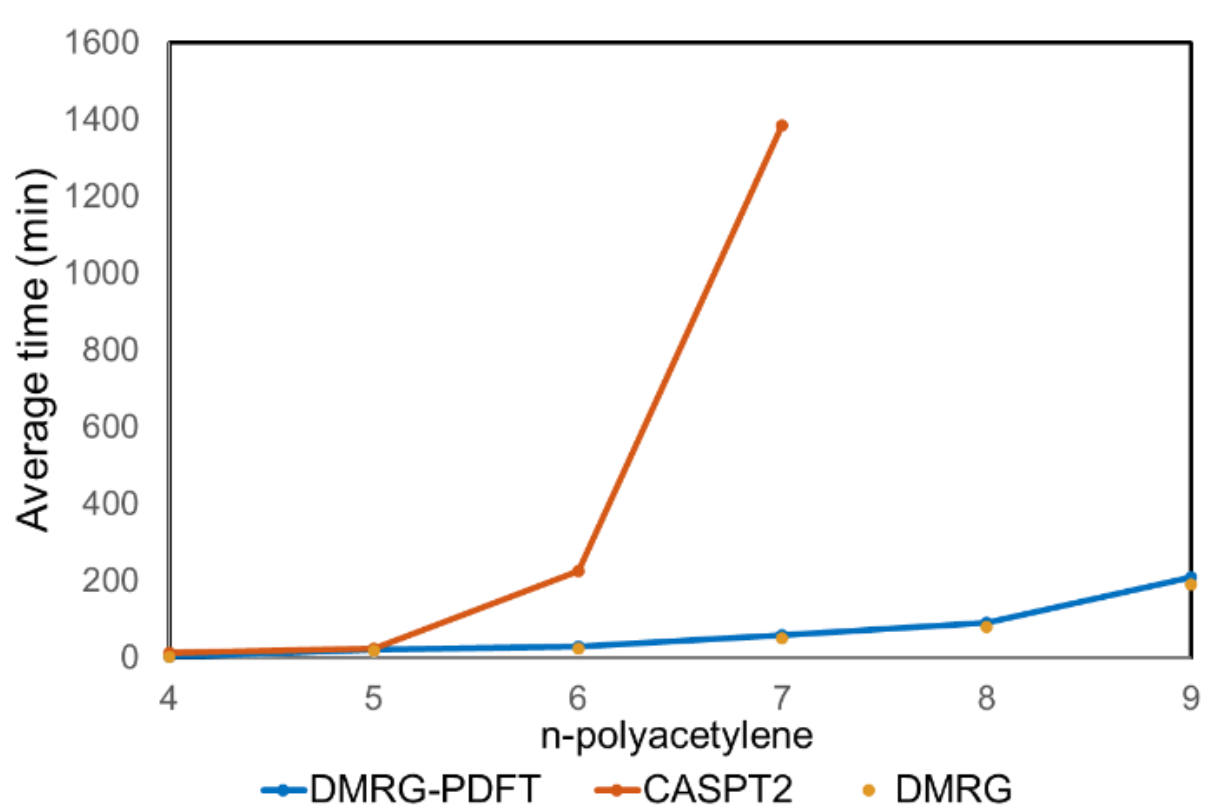

**Fig. 4**. Average compute time required for DMRG, DMRG-PDFT, and CASPT2 calculations on polyacetylenes with a single processor. The DMRG and DMRG-PDFT data are indistinguishable in the plot.

## 5 Conclusions

We have presented a new method, DMRG-PDFT, that combines the advantages of two approaches DMRG and MC-PDFT approaches. We used the resulting new method to calculate singlet–triplet gaps in polyacenes and polyacetylenes. The energy gaps calculated using DMRG were found to be close to the reference data, and we found that the DMRG-PDFT calculations are in most of the cases even more accurate than their bare DMRG counterparts. For polyacenes, singlet-triplet energies match reasonably well with the literature including GAS-PDFT values. For polyacetylene systems, we compared DMRG and DMRG-PDFT values with $M = 50$ to standard CASSCF and CAS-PDFT, and we found that DMRG-PDFT shows less dependence on the bond dimension than DMRG. Using the on-top density functional to add additional correlation energy to DMRG wave functions is shown to be a promising approach to study large systems at an affordable cost.

## Acknowledgment

This work was supported in part by Air Force Office of Scientific Research grant FA9550-16-1-0134. SK thanks Markus Reiher (ETH Zürich) for his continuous support.